# Graph Representation Learning Towards Patents Network Analysis

Mohammad Heydari
School of Industrial and Systems Engineering
Tarbiat Modares University
Tehran, Iran
m_heydari@modares.ac.ir

Babak Teimourpour*
School of Industrial and Systems Engineering
Tarbiat Modares University
Tehran, Iran
b.teimourpour@modares.ac.ir

*Abstract*— **Patent analysis has recently been recognized as a powerful technique for large companies in the world to lend them insight into the age of competition among various industries. This technique is considered a shortcut for developing countries since it can significantly accelerate their technology development. Therefore, as an inevitable process, patent analysis can be utilized to monitor rival companies and diverse industries. In this research, a graph representation learning approach employed to create, analyze and find similarities of the patents data registered in the Iranian Official Gazette. The patent records were scrapped and wrangled through the Iranian Official Gazette portal. Afterward, the key entities were extracted from the scrapped patents dataset to create the Iranian patents graph from scratch based on novel natural language processing and entity resolution technique. Finally, thanks to the utilization of novel graph algorithms and text mining methods, we identified new areas of industry and research from Iranian patent data, which can be used extensively to prevent duplicate patents, familiarity with similar and connected inventions, Awareness of legal entities supporting patents and knowledge of researchers and linked stakeholders in a particular research field.**

Keywords— Graph Representation Learning, Deep Learning, Patents Analysis, Graph Algorithms

## I. Introduction

Currently, the international economy depends on technological innovation. In the last two decades, there have been significant developments in the field of patent analytics. Patent application is a major side of protecting intellectual properties. Patent registry documents are a wide resource of technical and commercial knowledge. Therefore, patent analysis has been considered as a useful tool for managing research and development as well as technical and economic analysis. Patent's data is a complicated source for processing and gaining results albeit with a lot of human efforts and vast analysis time. There are many tools available to help patent experts and decision makers in patent analysis. As shown in previous research, patents data can be used to identify trends in industry as well as competitive power of enterprises or countries to track innovative activities. The most valuable benefits in patent networks analysis which can be noted include avoiding duplication in patent registry, innovators' familiarity with similar innovators and their mutual inventions, awareness of patent law enforcement agencies, and identification of relevant researchers in a particular research field. Consequently, patents data are analyzed by different techniques depending on the purpose pursued. Companies are interested in discovering latent patterns in patents for determination of patent innovation, identification of patent subjects, projection of technological innovations in a specific domain, industrial strategic planning, infringements detection in patent registry process, evaluation of patent quality for research purposes, road mapping industry, and recognition of industrial rivals. For patent data sources to be useful in decision making, the data must be precise, presented in a perceptible model, and delivered in the right manner. Patents, as the main sources of information, are significant to research and theoretical growth, serving as distinct information sources for technological information. Firms and individual inventors can achieve victory or lose essential income and market advantage if their modern designs inadvertently conflict or infringe the existing technology or if others misappropriate their pre-existing assertions. In the strictest sense, the oneness of an invention can be identified not by the occurrences of keywords and key phrases but by the inventive key findings or fresh proficiency [1][2][3].

## II. Background

There have been a few studies [4][5][6][7][8] on Iranian patenting works outside Iran, mostly conducted in US, yet until very recently, there had been no effort to carry out deep exploration into the patents registered in Iran Patent Office. [9] The main reason was that the Iran Patent Office did not have an official portal; therefore, not enough information or statistics could be obtained on patenting status in Iran. Iranian Official Gazette is an affiliate organization of Iranian judiciary in charge of publishing many kinds of public or legal notices, including acts of the parliament, laws and regulations, company registrations and deregistration, wills, trademark registrations, and patent registration. While many applicants were dissatisfied with the lengthy process of patent registration in Iran, Patent Office launched an e-filing service for patenting in July 2012. Ever since, applicants can fill all the required forms, pay the prescribed fees, and check the progress of their application through an online portal. This was a decisive step for the patent system to improve and expand its services in a large, geographically dispersed country like Iran. Iranian Official Gazette launched an online explorable database to prepare free of charge global access to information content of the patents registered in Iran. Iranian patents are granted for a maximum period of 20 years. The patent's owner shall have the exclusive rights of production, sale, and utilization of the subject of the patent. The latest studies investigated innovations in Iran by utilizing international databases such as USPTO, WIPO, and EPO. Unlike previous works, the most valuable point of this study is utilization of patents registered by the Iran Patent Office for the first time to the best of the authors' knowledge. Tehranchi provided an objective assessment of the Iranian American contributions to United States' science and technology as measured by their inventions registered with the United States Patent and Trademark Office (USPTO). The study is based on data collected by U.S. Census Bureau in 2000. Noruzi et al, mapped Iranian patents based on International Patent Classification (IPC) during 1976-2011. It also utilized non-national patents database such as USPTO, WIPO, and EPO. Madani et al, studied the evolution of patent mining. They applied bibliometrics analysis and keyword network analysis to 143 papers extracted from the 'Web of Science' database [10][11][12].

## III. OBJECTIVES

Mapping the patent-activity of a country based on IPC is the straightest method to calculate the technological specialization and technological scope of the country through distribution of its patents over different technological areas. The purpose of this study is to map the patents registered in Iranian Official Gazette Portal[1], known as the main patent registry base in Iran, in 2016–2019. The study attempts to address the three following questions:

- What are the most significant IPCs among the patented inventions?
- Which invention is like the patented invention?
- Who is the most potential innovator in each specific innovation field?

So, the study is to distinguish and discover the patterns in the innovation-activity in Iran, and accordingly shape a new landscape for future visionary and experimental research by design a novel recommender system in patents network. This study analyzes patent classification to explore the Iranian patent-activity and industrial development, examining Iran's patent areas in a long-term perspective. The patents are classified according to the International Patent Classification (IPC) format. IPC divides patentable technology into eight main categories as follows:

A. Human necessities, B. Performing operations; Transporting, C. Chemistry; Metallurgy, D. Textiles; Paper, E. Fixed construction, F. Mechanical engineering; Lighting; Heating; Weapons., G. Physics and H. Electricity. IPC is the main patent classification system in the world. Therefore, it served as the primary pattern for categorization of the patent subjects in this study. Each IPC has its corresponding technology classification. The IPC classification analysis helps researchers evaluate technology classification distribution of patents, assessing the overall technology trend of a country over different time periods. The inventors in this study were not limited to Iran; they were from different nationalities and countries among them Japan, China, North Korea (Asia), Netherlands, Germany, France (Europe), and USA.

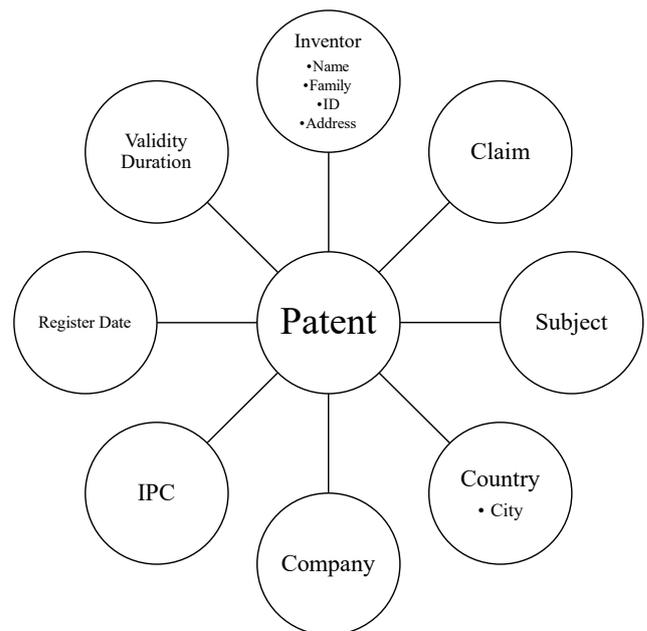

Figure 1 - Patent Entity Properties

## IV. RESEARCH METHODOLOGY

Research was carried out in the following brief steps:

A. Repository Selection
- Web scrapping by designing a web crawler to collect patents advertisement data.
- Data preprocessing and cleaning by utilizing Persian text processing techniques.
- Reshape data into the structured style.

B. Creation Of Patents Structural Database
- Extraction of necessary information about legal supporting institutions
- Extraction of necessary information about innovator (e.g., gender and nationality)
- Extraction of necessary information about patents (IPC, Subject, Innovators' names, and Owners)

---

[1] http://rrk.ir/News/NewsList.aspx

C. Graph Construction
- Apply graph deep learning algorithm on patents network to detect similar patents based on mutual IPC and IPC3 fields.
- Recommend similar nodes based on the node's properties similarity.
- Discovering various latent patterns in patents network.

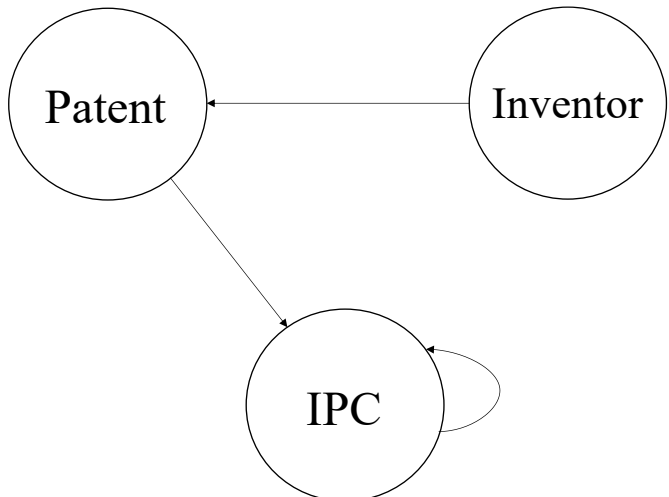

Figure 2 - Graph Schema

Each patent was registered with a unique ID is supported by an exclusive legal institution. Patents can contain mutual IPCs. The network consists of 6443 nodes and 8928 edges. Since the patents graph is a triplet graph with three key entities, the nodes refer to the patents, IPCs, and legal supporting institutions, and edges define the relationships between the node's connections to each other. The number of recognized innovators in our data set is 21465, who collaborated partially in innovation processes. Since the patents graph is a triplet graph with three key entities, nodes refer to the patents, IPCs, and legal supporting institutions, and edges define the relationships between the nodes' relations to each other.

## V. DATA

In this section, we talk about web crawler architecture to collect and patent data cleaning methods.

### A. Web Crawling

To implement the scrapping phase, web crawling technique was utilized to gather the desired data and create the patents dataset. A crawler was developed from scratch to fetch target data from the patents repository. It is worth mentioning that all the crawled data, including nodes and the relationships between them, are public data available to all users on the Iranian Official Gazette Portal.

### B. Data Processing

After scrapping data from the patents registry repository using the crawling technique, a specific regular expression was developed to identify the desired entities in a suitable Persian language style. Since the patents were recorded in the repository in Persian and this language poses some incompatibilities in text processing, a heavy preprocessing seemed vital to obtain the key entities in the standard format.

### C. Patents Data Extraction

As shown in the image below, the example of a patent advertisement is an HTML page encoded with the Unicode standard, which cannot be read by humans. To decode Unicode codes and convert them into Persian letters, the HTML2Text library has been used. It's a PHP library to convert HTML to formatted plain text.

Table 1 – A Patent Advertisement on Iranian Official Gazette

| مقدار | فیلد |
|---|---|
| تهران ۲۲۷۶۷ | شماره روزنامه |
| ۳۶ | شماره صفحه روزنامه |
| ۲/۳/۱۴۰۲ | تاریخ روزنامه |
| ۱۴۰۲۵۰۸۴۰۰۰۳۰۰۰۳۲۶ | شماره نامه اداره ثبت |
| ۳۱/۲/۱۴۰۲ | تاریخ نامه اداره ثبت |
| ۱۳۹۹۵۰۱۴۰۰۰۳۰۰۶۲۹۷ | شماره اظهارنامه اختراع |
| ۱۳۹۹/۰۷/۲۰ | مورخ |
| ۱۰۹۲۵۲ | شماره ثبت اختراع |
| ۱۴۰۲/۰۲/۳۱ | مورخ |
| آقای آیدین اشرفی بلگاباد | مالک اختراع |
| استان آذربایجان شرقی - شهرستان مراغه، ... | نشانی |
| ایرانی | تابعیت |
| آیدین اشرفی بلگاباد | نام مخترع |
| استان آذربایجان شرقی - شهرستان مراغه، ... | نشانی |
| سیستمی برای حفظ فاصله صفحه نمایش گوشی‌های هوشمند از صورت فرد استفاده‌کننده (کاربر) و اعلام هشدار به کاربر جهت تغییر موقعیت گوشی با استفاده از هوش مصنوعی | موضوع اختراع |
| H۰۴M ۱/۰۰ | طبقه بندی بین المللی |
| از تاریخ ۱۳۹۹/۰۷/۲۰ الی ۱۴۱۹/۰۷/۲۰ به مدت ۲۰ سال می باشد. | مدت حمایت |

## VI. NETWORK SCIENCE APPROACHES

The outcome of our study is divided into different sections, so it could help design a technology roadmap, increase products reusability, and save time as the most asset.

### A. Graph Structural Information

The table below shows the structural information of the graph. Like any other object in the real world, each network incorporates some features which can serve a strategic role to give the observers more insight. Based on the distribution degree of the whole graph, a small number of nodes have degrees in the range of 50–436. Most of the nodes have degrees in a range under 50. Almost %10 of the nodes (650 items to be exact) are isolated as they do not play a significant role. It can be concluded that this small percentage of nodes is not mutual with other nodes in IPC. Their research fields could be too old or so novel that they have not yet formed a significant community or drawn interest. Since graph distribution degree follows power law type, it can be mentioned that the network is scale-free. As mentioned, there are a small number of nodes with high range of degrees and many nodes with a small range of degrees. The average degree of the nodes is 2.07. The average path length between the nodes is 5.08. The modularity value is 0.771.

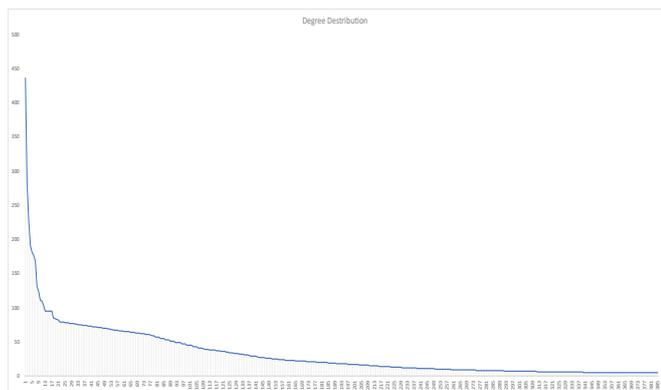

Figure 3 - Patens Graph Degree Distribution

## B. Network Centralities

In this part, we demonstrate top patent graph nodes based on degree centrality.

Table 2 - Top Patents IPC Classes based on Frequency.

| Rank | IPC | Frequency | Percentage |
|---|---|---|---|
| 1 | A | 1653 | 28.8 |
| 2 | B | 1074 | 18.62 |
| 3 | C | 929 | 16.12 |
| 4 | F | 716 | 12.42 |
| 5 | G | 702 | 12.18 |
| 6 | H | 334 | 5.79 |
| 7 | E | 270 | 4.68 |
| 8 | D | 84 | 1.45 |

Table 3 - Top Patents IPC Subclasses based on Graph Degree Centrality

| IPC | Class | Subclass | Deg Cent |
|---|---|---|---|
| A61K | A | Preparations for Medical, Dental | 436 |
| B82Y | B | Applications of Nanostructures | 282 |
| A61B | A | Diagnosis; Surgery; Identification | 226 |
| B01J | B | Chemical or Physical Processes | 191 |
| G05B | G | Control or Regulate Systems in General | 181 |
| F01B | F | Positive-Displacement Type, Steam Engines | 177 |
| G01N | G | Investigating or Analyzing Materials by Determining Their Chemical or Physical Properties | 168 |
| A61F | A | Filters Implantable into Blood Vessel, Prostheses | 131 |
| B01D | B | separating solids from solids by wet methods | 124 |
| C02F | C | processes for making harmful chemical substances harmless | 111 |

The table entails the most precious IPC extracted from the patent network, presented based on frequency and listed in the table according to ranking. It is obvious that most of the patents heavily deal with Human Necessities, Performing Operations, and Chemistry.

## C. Patents keyword Extraction

In this part, visualization of the patents network and three different snapshots of the graph are shown. The points in the images are hubs distributions. Most key communications in the graph initiated through the highlighted hubs, which were elaborated in the Results section.

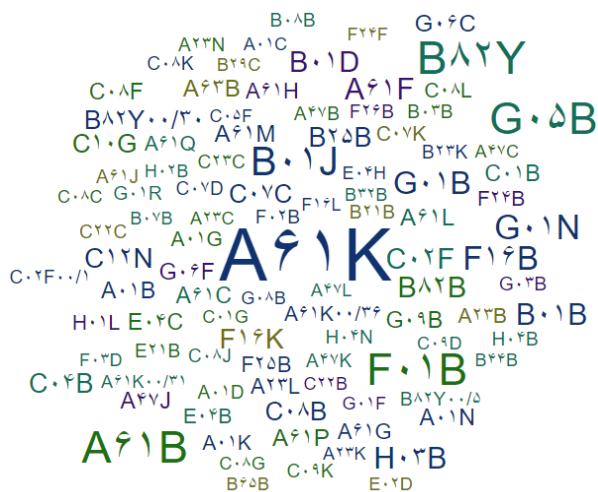

Figure 4 - Visualization of Most frequented IPC's

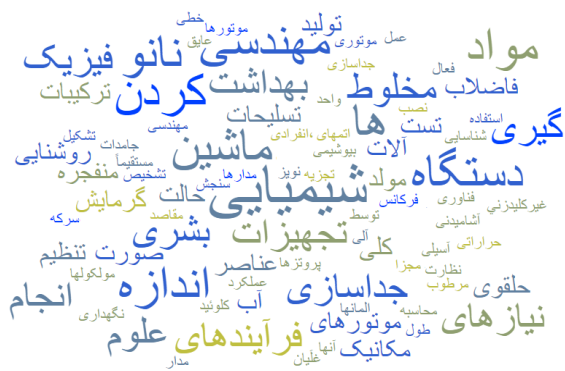

Figure 5 – Top Keyword Extracted from Patents Different Topics

In the following diagrams, some of the most important results about the patents data are listed. As is shown, in the various types of Patents Legal Supporting Centers, the companies are highlighted according to their distinguished contribution in the patent's registry process with 928 cases.

## D. Patents Graph Visualization

The idea was to find which edges in a network occur most frequently between other pairs of nodes by finding edges betweenness centrality. The edges joining communities are then expected to have a high edge betweenness. The underlying community structure of the network will be much more fine-grained once the edges with the highest betweenness are eliminated which means that communities will be much easier to spot. In the following visualization, which is based on Girvan-Newman community detection algorithm[13], each color represents a specific community of graph. Girvan-Newman method [14] is one of the classic community clustering techniques. By using the algorithm, we can separate the network into communities, and the community detection can be used as a good start of data preprocessing. It will remove the edges with the largest edge betweenness in every iteration. It relies on the iterative elimination of edges that have the highest number of shortest paths between nodes passing through them. By removing edges from the graph one-by-one, the network breaks down into smaller pieces, so-called communities. The algorithm was introduced by Michelle Girvan and Mark Newman.

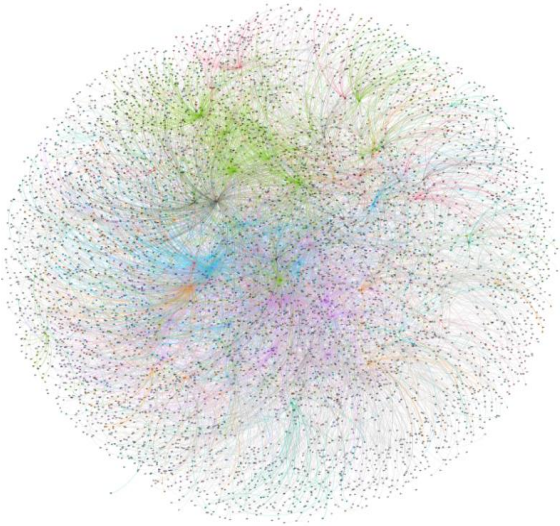

Figure 6 - Aach color refers to a unique community of Patents.

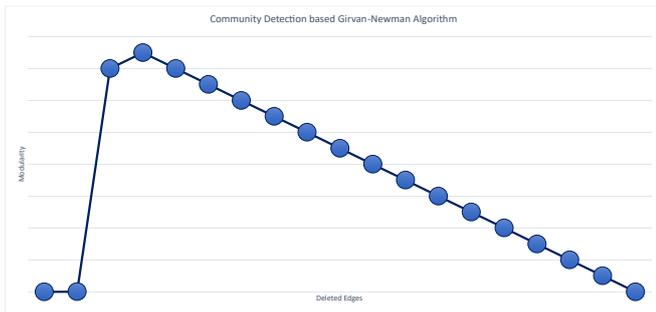

Figure 7 – Community Detection based Girvan-Newman Algorithm

Based on report, Number of communities are 147 and Maximum found modularity is 0.7719164.

In the following diagrams, some of the most important results about the patents data are listed. As is shown, in the various types of Patents Legal Supporting Centers, the companies are highlighted according to their distinguished contribution in the patent's registry process with 928 cases.

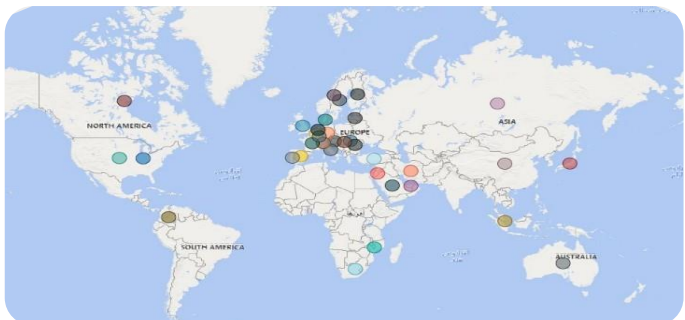

Figure 8 - Geographical Distribution of Innovators' Nationalities

As it is shown in the above visualization, it demonstrates the geographical distribution of innovators' nationalities around the world. This visualization was produced by Power BI software. Since foreign investors can register their patents in Iranian Official Gazette Portal, there are various nationalities among the innovators. After Iran, Europe, East Asia, and America had the most active collaboration, respectively. Netherlands, Germany, France, and Italy, representing Europe, with an overall of 163 participations, had the second biggest participation after Iran. Japan, China, and North Korea, representing Asia are the third major collaborator after Europe, with 147 participations. Finally, USA stood as the last with 23 cases of collaboration.

## VII. Graph Representation Learning

In this section, the representation of graph results based on DeepWalk[15], Node2Vec[16] inspired by Word2Vec[17], LINE and SDNE is shown. After applying multiple random walks and obtaining the latent vectors of the nodes from the heterogeneous graph, they can be displayed in a two-dimensional space using the T-SNE algorithm, which is a dimensionality reduction algorithm. The images below are the two-dimensional representation of the features of the nodes in the two-dimensional space. The purpose is to embed the patents and IPCs by the following steps. (1) Nodes will represent patents and IPCs. (2) Each patent is connected to its specific IPC given that a field named IPC3 was created, referring to mutual IPCs of a patent. It shows that a patent can contain more than one IPC and connect several patents. (3) Node2vec is applied to the resulting graphs. At the end, the similarity between the different nodes is inspected. It is expected that the most similar nodes to a patent be the ones with mutual IPCs and IPC3. As mentioned above, IPC3 refers to mutual IPCs of a patent. The art of algorithm is automatic node embedding learning. The patent ID fields were given as the input to the model to feed it and return the most similar patents with mutual IPC and IPC3. IPCs acted as the hubs in the network to facilitates patent relations to each other.

### A. Deep Walk

Table 4 - DeepWalk Algorithm Parameters

| Parameters | Values |
| --- | --- |
| Walk Length | 10 |
| Number of Walks | 80 |

### B. Node2Vec

Table 5 – Node2Vec Algorithm Parameters

| Parameters | Values |
| --- | --- |
| Dimension | 64 |
| Walk Length | 30 |
| Number of Walks | 200 |
| Window Size | 5 |
| Epochs | 3 |

### C. LINE

Table 6 - LINE Algorithm Parameters

| Parameters | Values |
| --- | --- |
| Representation Size | 128 |
| Order | 2 |
| Batch Size | 1024 |
| Epochs | 50 |

### D. SNDE

Table 7 – SNDE Algorithm Parameters

| Parameters | Values |
| --- | --- |
| Hidden Size | 256, 128 |
| Batch Size | 3000 |
| Epochs | 40 |

One of the common methods of checking graph properties is to use the representation method. Many common methods in machine learning, tests and statistical inferences are defined in vector spaces, that is why methods that take a graph as input and output a representation of the vertices or edges of this graph in a vector space, such as vertices, have become common. Similar or neighbors have a small Euclidean distance. By using this algorithm automatically, we learn the characteristics of each node in the communication graph and create a latent vector for them. Then we will predict the communication using the classification algorithm. This algorithm helps us to identify nodes with the highest degree of similarity. The criterion of similarity is the maximum sharing of inventions based on the international classification codes of inventions. So obviously, the output of the algorithm implementation on the patents network graph are the most similar nodes (inventions) to each other. To implement the random walk algorithm in the Node2Vec core, the model parameters must be adjusted. Next, the process of generating random steps is simulated and the model is trained on the graph data set. The input of the model is the patent identifiers, and the output of the model is semantically similar nodes that have international patent classification codes in common. The important point here is that we did not transfer any information about the structural characteristics of the graph nodes to the model in advance, and the model learned the representation automatically. After the end of the model training process, the most similar nodes are identified. Our input for testing the recommender system is one of the nodes of the graph that refers to an invention, and the result of the recommender system is similar inventions (similar nodes) along with the similarity score of each invention.

## VIII. CONCLUSION

Patent analysis plays a pivotal role in understanding technological advancements and competitive landscapes across industries. In this context, recommender systems within patent networks offer indispensable advantages, from preventing duplicate patents to fostering knowledge sharing among investors and institutions. The uniqueness of our approach lies in its application to Iranian patent data, opening new vistas for minimizing redundancy, enhancing familiarity with related inventions, understanding legal facets, and connecting researchers and stakeholders within specific research domains. The study recognizes the crucial role of patent analysis in gaining insights into technological advancements and competitive landscapes in various industries. It emphasizes the importance of recommender systems in patent networks to prevent duplicate patents, facilitate knowledge sharing, and support inventors and institutions. The paper's main contributions include the utilization of graph representation learning approaches, specifically graph deep learning algorithms, to uncover latent patterns in patent networks and identify similar nodes. By analyzing Iranian patent data, the study opens new avenues for preventing duplicate patents, promoting familiarity with related inventions, understanding legal aspects, and connecting researchers and stakeholders within specific research fields. The research methodology involved web scraping and data preprocessing to create a structured patent database. Graph construction and deep learning algorithms were applied to identify patent similarities, legal institutions, innovators, and patent subjects. The paper also provides insights into patent classifications, centralities, and community structures within the network. Through visualization and representation learning techniques such as DeepWalk, Node2Vec, LINE, and SDNE, the study offers a comprehensive understanding of patent relationships and similarities. These methods enable the automatic learning of node characteristics and the prediction of similar patents based on international classification codes. Overall, this paper's findings have significant implications for the Iranian patent ecosystem. The recommender system developed in this research can facilitate innovation, reduce redundancy, and enhance collaboration among investors and institutions. It provides a valuable tool for patent analysts, researchers, and decision-makers in navigating the complex landscape of patent data and fostering technological advancement.

**Appendices**

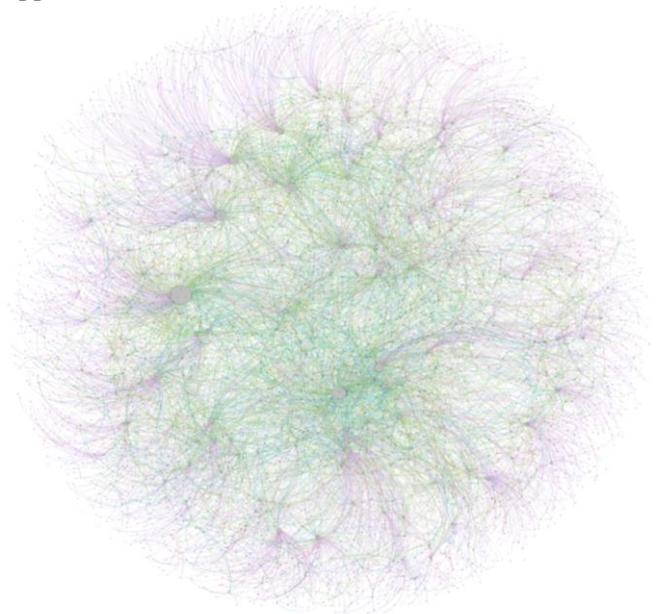

Figure 9 - Graph Visualization based on Degree Centrality

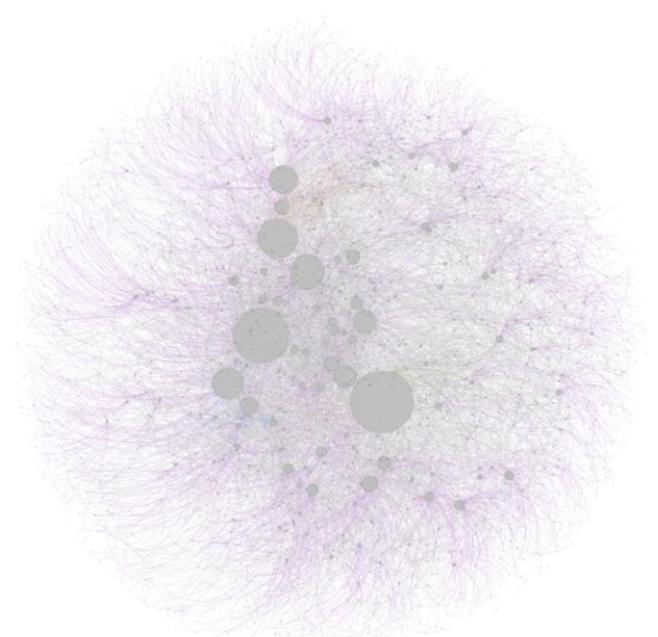

Figure 10 - Graph Visualization based on Betweenness Centrality

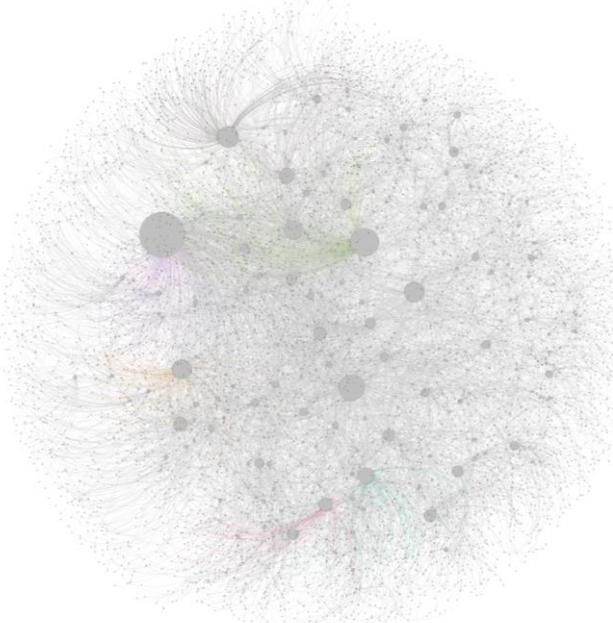

Figure 11 - Graph Visualization based on PageRank Centrality

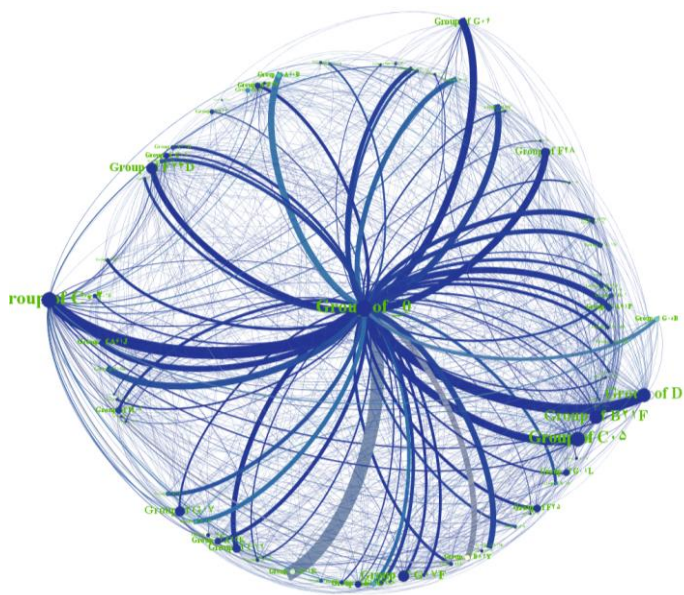

Figure 12 - Visualization of the associations identified in the graph (interweaving and sharing of IPC codes)

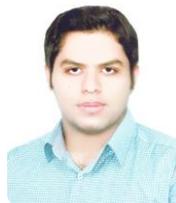

Mohammad Heydari received his B.Sc. degree in Computer Software Engineering, Technical and Vocational University of Tehran, Iran and received his M.Sc. degree in Information Technology Engineering, Tarbiat Modares University of Tehran, Iran. His research interests include Data Science in Healthcare, Intersection of Machine Learning and Network Science, Big Data Analytics, and Bioinformatics. He's the Founder of Data Science School, An AI Tutorial Platform with a creative license from Vice-Presidency for Science and Technology of Iran.

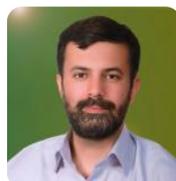

Babak Teimourpour received his B.Sc. degree in Industrial Engineering, Sharif University, Tehran, Iran in 1996 and received his M.Sc. degree in SocioEconomic Systems Engineering, Institute for Research on Planning and Development, Tehran, Iran in 1998. Also, he received his Ph.D. in Industrial Engineering at Tarbiat Modares University, Tehran, Iran in 2010. His research interests include Data Mining, Social network analysis and Bioinformatics.